\documentclass[12pt,a4paper]{article}
\usepackage[utf8]{inputenc}
\usepackage{graphicx}
\usepackage{amsmath}
\usepackage{booktabs}
\usepackage{array}
\usepackage{caption}
\usepackage{subcaption}
\usepackage{setspace}
\usepackage{geometry}
\usepackage{hyperref}
\usepackage{natbib}
\usepackage{multirow}
\usepackage{amsfonts}
\usepackage{amssymb}

\title{AI-Enhanced TOE Framework for Sustainable Industrial Performance in Fragile and Transforming Economies: Evidence from Yemen and Saudi Arabia}
\author{
    Shaima Farhan\textsuperscript{a}, 
    Dong Yu\textsuperscript{a}\textsuperscript{*}, 
    Amirhossein Karamoozian\textsuperscript{c}, 
    Ali Al-shawafi\textsuperscript{d}, \\
    Amar N. Alsheavi\textsuperscript{f}
}

\date{}

\begin{document}

\maketitle

\begin{center}
    \begin{tabular}{ll}
        \textsuperscript{a} & School of Management, University of Science and Technology of China, \\
        & Hefei, Anhui, 230027, China \\
        \textsuperscript{c} & School of Management, University of Science and Technology of China, \\
        & Hefei, Anhui, 230027, China \\
        \textsuperscript{d} & School of Civil Engineering, Tianjin University, Tianjin, 300350, China \\
        \textsuperscript{f} & School of Computer Science and Technology, University of Science and  Technology of China, \\
        & Hefei, Anhui, 230027, China \\
    \end{tabular}
    
    \vspace{0.5cm}
    
    \textsuperscript{*}\textbf{Corresponding author:} Dong Yu, School of Management, University of Science and Technology of China, Hefei, Anhui, 230027, China. E-mail: ydong@ustc.edu.cn
\end{center}

\begin{abstract}

Using an integrated framework rooted in the TOE model enhanced with AI, this study looks at ways to improve industrial performance and environmental sustainability in fragile and rapidly transforming contexts such as those found in Yemen and Saudi Arabia. Data for the research are field-based and were obtained from a total of 600 SMEs operating in both countries. Based on the questionnaires' response by 294 managers, results from the partial least square structural equation modeling (PLS-SEM) have indicated significant positive effects of AI-TOE on environmental performance ($\beta = 0.487$) and manufacturing performance ($\beta = 0.759$). Results indicate that AI acts as a transformative force, though its impact differs based on the maturity of infrastructure and the organizational readiness. The Saudi SMEs gain from the institutional support and advanced technologies, while those in Yemen are dependent on the low-cost solution of AI and organizational flexibility to accept the strength of structural challenges. PLS-SEM analysis of the study showed that integrating AI into the TOE dimensions further accelerates the operational efficiency in order to support environmental performance.Industrial performance was found as a very important mediator in this relationship. This study responds to the call for digital transformation literature by providing an actionable framework of AI adoption in resource constrained environments. This study offers insights that might guide policymakers and organizations toward more resilient and sustainable operational strategies. These findings provide invaluable guidance for engineering managers within the context of negotiating digital transformation and sustainability trade offs in fragile and resource-constrained contexts.

\end{abstract}

\section{Introduction}
AI-driven digital transformations have positioned industries in developing economies, such as Yemen and Saudi Arabia, to advance beyond their existing levels \citep{Fossung2024}. Recent studies on AI adoption presented the fragile economy technology adoption, or FETA, framework, which extended the traditional TOE model by incorporating the capabilities of AI into all three dimensions to address unique challenges existing in resource-constrained environments: survival-oriented innovation, organizational flexibility, and institutional adaptation through AI integration in order to evaluate performance implications for SMEs operating in resource-constrained environments  \citep{Santhi2023}.
Unlike the TOE studies conducted in stable settings \citep{NDri2024, Satyro2024}, this research looks into technological leapfrogging in fragile contexts-that is, when countries leapfrog certain developmental stages by leveraging advanced technologies \citep{Christensen2019}. Previous literature indicated that AI improves manufacturing and environmental performance along with green innovation \citep{Akram2024, AlKoliby2025}. This research contributes to the literature by indicating how AI can have a differential impact on both industrial and environmental performance in fragile economies. The Granger comparison analysis between Yemen and Saudi Arabia informs SMEs in fragile economies on how to improve their performance and contribute towards environmental concerns through judicious use of resources and adaptations in innovation practices.

This framework offers a contextualized TOE that shows AI as an innovative strategy propelled by fragility and institutional transformation. We focus on the case of artificial intelligence, because it is arguably the most transformative technological force that can enable SMEs in fragile economies to leap beyond conventional developmental stages while simultaneously tackling a series of operational and environmental challenges \citep{Jan2023, Tseng2024}. This study will uniquely explore the issue of how AI acts as a strategic enabler within the continuum of fragility and stability.

These specific implications of AI in our research context are multilayered: First, it enables "survival-oriented innovation" in highly fragile contexts such as Yemen-which ranked first in the world on the Fragile States Index-in which organizations adopt low-cost AI solutions to continue operations in a context of institutional collapse; Second, it responds to "institutional lag" in fast-transforming economies such as Saudi Arabia, which is showing rapid transformation through Vision 2030 \citep{SaudiArabia2016}, and in which AI capabilities are enabling organizations to respond to policy changes more swiftly than traditional mechanisms of institutional change \citep{North1990}; Third, it provides a technological bridge by which both contexts can achieve concurrent improvements in manufacturing performance and environmental sustainability; And, fourth, predictive insights regarding AI-driven environmental transitions, including the "AI Environmental Kuznets Curve" phenomenon documented in Saudi Arabia, extending Porter and van der Linde's hypothesis on environmental competitiveness to AI application \citep{Porter2002}.

Against this backdrop, AI becomes the strategic technological intervention to reduce the gap between institutional fragility and economic resilience ,. The study has important strategic imperatives for responsible AI investments in contexts faced with institutional incapacities, human resource constraints, and ecological imperatives amidst turbulent conditions ,. In fact, these findings inform decision frameworks that would be employed to post-conflict reconstructions, climate-exposed economies, and rapid transformation initiatives worldwide, as argued in contemporary debates on digital resilience scholarship,.

This therefore represents the first holistic AI-integrated TOE framework designed for fragile and transforming economies, hence filling a critical lacuna in the existing literature which has so far been predominantly set within stable economic contexts. Whereas past studies have explored AI adoption in isolation, this study is unique in showing how AI works to be a transformative bridge across the fragility-stability continuum and enable betterment in manufacturing efficiency and environmental sustainability simultaneously. Theoretically, it extends the traditional TOE model by integrating survival-oriented innovation strategies and introduces the new concept of "institutional leapfrogging" in AI adoption patterns across divergent economic contexts.

\section{Literature Review}

AI is the most disruptive technological paradigm of the 21st century in terms of the radical change it makes in organizational capabilities and the competitive landscape for different industries. A number of studies from technological forecasting and social change have illustrated that AI technologies offer unparalleled opportunities to advance organizational performance. For instance, \citet{Bonab2023} investigated how generative AI and conversational AI technologies have been transforming business practices across different industries. The adoption of AI is a complex issue and therefore warrants a robust theoretical framework that can conceptualize its multidimensional nature. It has shown how technological capability, organizational readiness, and environmental pressures together create an outcome on digital transformation. \citet{Chapman2022} established that small and medium enterprises exhibit different patterns of digital transformation marked by flexibility, speed of decision-making, and novel resource utilisation strategies. Specifically, the findings have brought out the critical aspect of the role SMEs play in demonstrating that small businesses are indeed very significant in facilitating the diffusion in developing economies. The areas where AI-powered procurement and supply chain systems bring about significant improvements in operational performance include predictive analytics, automated decisionmaking, and smart resource optimization. This study identified how machine learning algorithms serve to attain significant improvements in operational efficiencies and competitive advantages.  Further, it was also found that heavy organizational context influences are linked with the strategic direction behind the implementation of AI. In linking AI adoption with environmental sustainability, it can be said that intelligent manufacturing systems contribute both to operational efficiency and to improvement in environmental performance. The technological dimension within AI integrated TOE frameworks encompasses not just conventional factors of technology adoption but also the requirement laid down by AI. A recent study by \citet{Muhammad2025} established that analytics capability, technological infrastructure, and integration complexity are major determinants of the successful digital transformation of any organization. Firms with sound technological readiness attain better AI adoption performance across various contexts.\citet{Huang2021} have presented the study on AI implementation Patterns. According to \citet{Nouira2022}, digital transformation outcomes and competitive advantages are significantly influenced by organizational capabilities such as learning capacity, technological sophistication, and the ability to manage changes.

\section{Hypotheses}

\subsection{AI-Supported TOE and Environmental Performance}

Indeed, the integration of AI into the TOE framework provides encouraging results related to the improvement of environmental performance in different organizational contexts. For this context, environmental performance means the minimisation of adverse impacts like emissions, generation of wastes and avoidable resource use that comply with legal requirements. Several works identified artificial intelligence as constructive in its role. For example, \citet{Pandey2024} found industrial operations to be less resource intensive and be more efficient due to AI applications. According to \citet{Graef2021}, AI systems helped reduce overproduction-based emissions by optimizing production quantity. Similarly, \citet{FossoWamba2024} found that AI-driven systems minimized hazardous wastes and were able to ensure compliance with the environment through continuous adjustment in the system, etc. Similarly, \citet{Centobelli2019} identified that innovative manufacturing strategies resulted in waste reduction; improvement in the quality and efficiency of production contributed to it. Again, \citet{Cannas2024} established that the AI-driven system contributed to improving competitive performance due to improvement in delivery timeline, shrinking of expenses and improvement in product quality.

Apart from this, lax or non-existent regulation has also formed another big stumbling block in the way to this area. In this respect, an organization needs to build and invest in technical and managerial competencies for AI \citet{DeGiovanni2021}. However, partnering with technology providers who can enable opportunities leading to long-term gains is required. Similarly, well-framed regulations that are aligned with environmental concerns could be of crucial help. This collaborative effort of academia, industry, and government can further accelerate innovative solution creation in respect of sustainability. Based on this review, one can hypothesize that :\\

H1: AI-supported TOE considerably and positively influences environmental performance.\\

H2: AI-supported TOE contributes positively to environmental performance as a result of improvements in manufacturing efficiency. \\
H3: Manufacturing performance has a positive influence on environmental performance.\\
H4: The potential of AI-supported TOE applies directly to manufacturing performance.\\

\begin{table}[htbp]
\centering
\caption{Summary of studies on AI and sustainability in SMEs (Part 1)}
\label{tab:litreview1}
\scriptsize
\begin{tabular}{p{2cm}p{2cm}p{2cm}p{3cm}p{3cm}}
\toprule
\textbf{Study} & \textbf{Focus Area} & \textbf{Context / Sample} & \textbf{Key Findings} & \textbf{Identified Gaps / Limitations} \\
\midrule
\citet{Pandey2024} & AI in reducing resource consumption & Industrial operations & AI contributes to resource efficiency and higher productivity & Lacks a long-term sustainability perspective and neglects manufacturing improvement; no attention to weak infrastructure environments. \\
\citet{FossoWamba2024} & AI in reducing hazardous waste and ensuring environmental compliance & Global case studies & AI helps minimize hazardous waste and enhance environmental compliance & Overreliance on AI may misalign with sustainability goals, as it lacks a focus on complex structural environments and the interplay between environmental and manufacturing dynamics. \\
\citet{AlKoliby2025} & Leadership, strategy, and AI in driving green innovation & 219 manufacturing SMEs in Malaysia & GKOL improves GI directly and indirectly via ES; GAIC strengthens GKOL's impact on GI & Limited generalizability; cross-sectional design; single-respondent bias; lacks details on AI types. \\
\citet{Sharma2022} & AI in environmental management & Indian SMEs & AI improves environmental management and supplier selection & It focuses on management without addressing broader institutional or manufacturing dimensions, with no attention to fragile environments. \\
\citet{Yang2022} & AI in improving manufacturing productivity & Global manufacturing firms & AI increases productivity while meeting environmental targets & Emphasis on large enterprises without considering resource-limited settings; lacks insight into contexts like Yemen. \\
\citet{Agrawal2023} & AI-driven efficiency in SME resource use & Indian SMEs & AI adoption improves productivity by optimizing resource use & It focuses solely on Indian SMEs and does not explore the challenges in other developing regions, such as Yemen. \\
\citet{Matin2023} & AI with IoT in manufacturing & 120 European companies & AI combined with IoT improves manufacturing and energy efficiency & Limited to European firms and lacking an emphasis on environmental sustainability, it does not explore developing country settings. \\
\citet{Liu2023} & AI for environmental innovation and carbon emission reduction & Various international firms & AI promotes green innovation and lowers carbon emissions & Focuses too narrowly on emissions and energy, overlooking broader sustainability goals. \\
\citet{Mumali2022} & AI in predictive maintenance & Manufacturing operations & AI-driven maintenance reduces costs and downtime & Does not explore barriers specific to small firms in adopting predictive technologies. \\
\citet{Cannas2024} & AI in enhancing competitive performance & SMEs & AI improves delivery, reduces costs, and enhances product quality & Lacks integration of AI with organizational capabilities in SMEs. \\
\citet{Gupta2022} & AI in production planning and forecasting & SMEs & AI enhances planning, resource allocation, and forecasting & Insufficient examination of institutional and infrastructure-related barriers. \\
\citet{DeGiovanni2021} & AI in operational efficiency and environmental goals & SMEs & AI supports demand forecasting and resource management & Neglects AI's organizational alignment and integration with operational systems. \\
\bottomrule
\end{tabular}
\end{table}

\begin{table}[htbp]
\centering
\caption{Summary of studies on AI and sustainability in SMEs (Part 2: Current Study)}
\label{tab:litreview2}
\scriptsize
\begin{tabular}{p{2cm}p{2cm}p{2cm}p{3cm}p{3cm}}
\toprule
\textbf{Study} & \textbf{Focus Area} & \textbf{Context / Sample} & \textbf{Key Findings} & \textbf{Identified Gaps / Limitations} \\
\midrule
\textbf{This Study (2025)} & AI impact on environmental and manufacturing performance in SMEs in Yemen and Saudi Arabia & SMEs in Yemen and Saudi Arabia & AI significantly improves environmental and manufacturing outcomes in challenging settings & It addresses infrastructure gaps in Yemen and policy misalignments in Saudi Arabia, and demonstrates how geographic and cultural differences impact the effectiveness of AI in enhancing environmental and manufacturing performance. It presents a unified strategic framework rather than treating the two domains separately. \\
\bottomrule
\end{tabular}
\end{table}
\begin{figure}[htbp]
\centering
\includegraphics[width=0.8\textwidth]{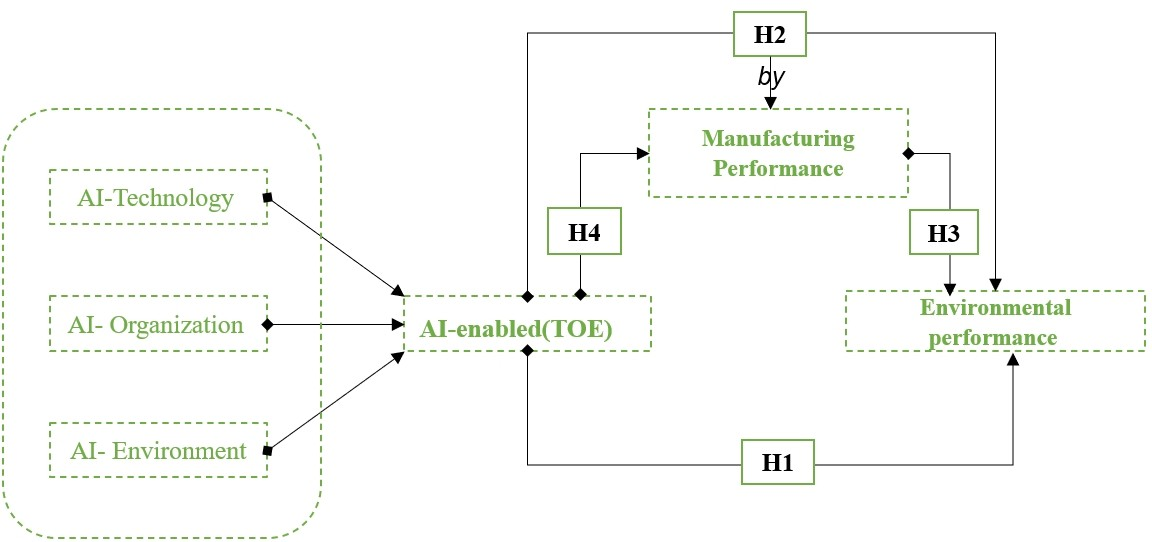}
\caption{Research model}
\label{fig:model1}
\end{figure}

\begin{figure}[htbp]
\centering
\includegraphics[width=0.6\textwidth]{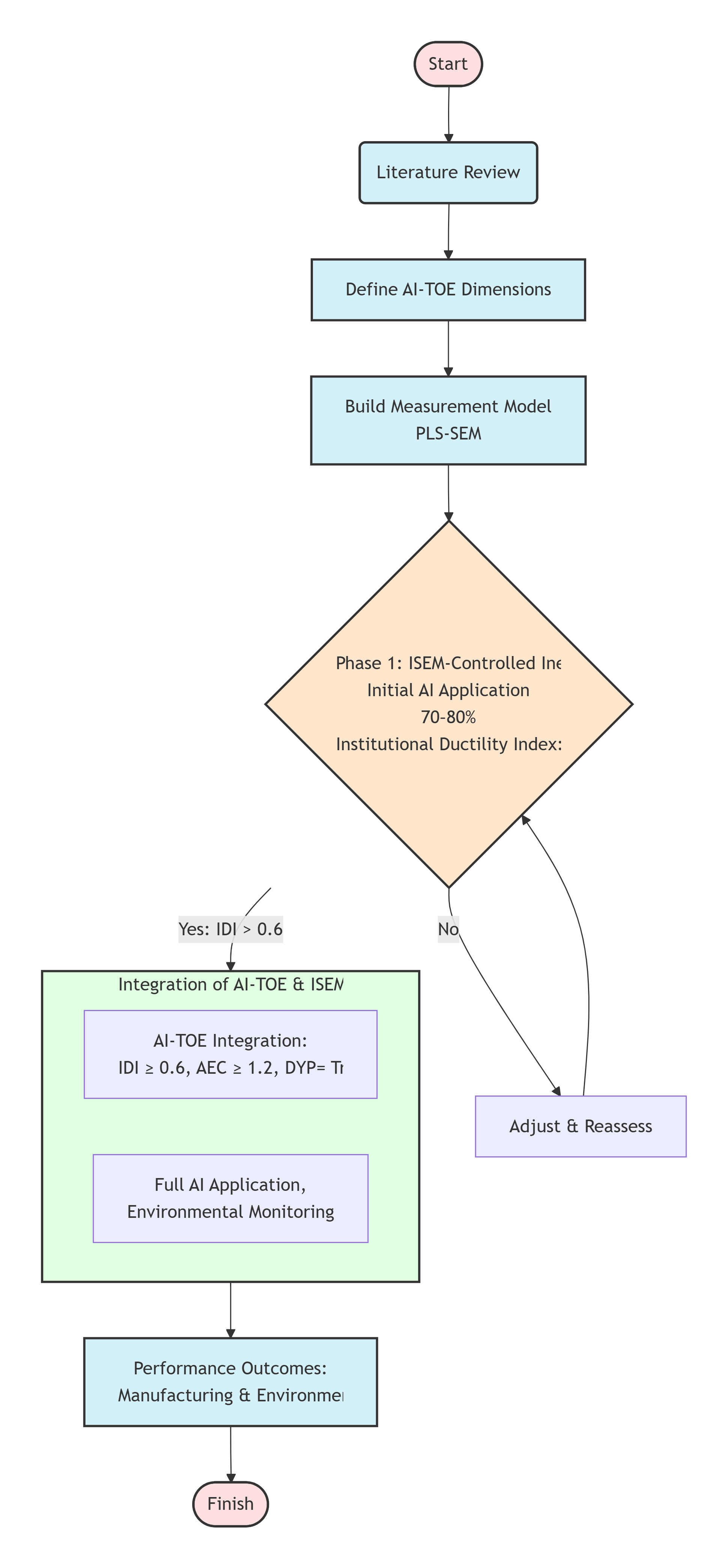}
\caption{General Framework for AI-TOE Integration}
\label{fig:model2}
\end{figure}

\section{Research Methodology}

\subsection{Sample and Data Collection}

These questions have been designed in light of observations and interviews previously conducted among SMEs in Yemen and Saudi Arabia. It came out during the interviews that companies in Saudi Arabia enjoy strong institutional stability along with advanced digital infrastructure and, as such, are capable of applying AI technologies like predictive maintenance systems, cloud-based ERP platforms, and data analytics tools to improve operational efficiency of organizations. Companies in Yemen, despite low conditions of infrastructures, reported use of simpler forms of AI solutions, like chatbots and inventory forecasting tools, which were more appropriate for their technological environments and business conditions prevailing in the locality. This has resulted in capturing an understanding of the technological development of the organizations in question and the challenges faced by them.

This gives sufficient geographic diversity and hence makes our sample representative of the population. A standard procedure was followed in developing the questionnaire of this study before data collection. First, we borrowed from the prior literature the measurement scales of the relevant constructs that were in existence and adapted them for use within the context of the AI-TOE. Since our sample of companies was in Yemen and Saudi Arabia, the original survey items in English were first translated into Arabic and then back-translated for accuracy and consistency. Third, we invited four Yemeni and Saudi researchers specializing in AI to review the measurement scales. These were helpful in pointing out some problems caused by the wording of certain questions. Finally, a pilot test was conducted in 30 companies each from Yemen and Saudi Arabia, whose responses were excluded from our final sample. The comments and suggestions furnished by the participants helped make many improvements in the clarity and relevance of the questionnaire. A questionnaire was developed based on this, with a clear purpose, completeness of contents, and consistency of structure.

Next, the process of data collection was explained in detail as follows: From the National Bureau of Statistics in Yemen and Saudi Arabia, we compiled a list of 1,006 firms distributed across urban and rural areas in Yemen and Saudi Arabia to understand technology diffusion across different regions. This may have had an impact on the findings, since one would expect the level of technological presence to vary across urban and rural areas. This was done as follows: 300 firms in Yemen, divided between urban areas (Sana'a, Aden, Mukalla) and rural areas (Hodeidah, Raymah, Al Mahwit), and 300 firms in Saudi Arabia, divided between urban areas (Riyadh, Jeddah, Dammam) and rural areas (Qassim, Taif, Najran). Firms in the former were included and selected based on their level of technological development. Specifically, six interviews were conducted with CEOs. Because CEOs are in the best position to identify suitable company employees to participate in the survey, they were contacted by telephone or email. Qualified participants recommended by the CEOs were then approached. What is more, in order to avoid cross-method bias, we divided this research into two parts and surveyed each candidate one month apart. In total, 1,006 SMEs were contacted in Yemen and Saudi Arabia. After two waves of data collection, 294 complete responses were retrieved. The response rate comes to 29.2 \%, with a balanced split between Yemen (n=147; 50\%) and Saudi Arabia (n=147; 50 \%). Following the protocols suggested by \citet{Dillman2014}, non-response bias was detected by performing chi-square analysis based on firm size and industry distribution between the respondents and the target population ($\chi^2 = 3.21, p > 0.05$). Table \ref{tab:demographics} summarizes the key characteristics of the study sample and accurately reflects the demographic distribution of the respondents. Data Analysis Standard method bias was used through IBM SPSS version 20 and Partial Least Squares (PLS) regression analysis in order to let both the measurement and structural models be estimated jointly. In that respect, SmartPLS version 3.2.7 was used for conducting the PLS estimation effectively. Common Method Bias and Multicollinearity Testing The first factor accounted for 35\% of the variance. The test by Harman suggests that the standard method bias is a problem if it is above 50\%. In this study, VIF values range between 1.170 and 2.595. Since multicollinearity is not found to be a problem if the VIF is less than 10, neither common method bias nor multicollinearity is a serious problem in this study.

\subsection{Country Selection Rationale}
This research applies a strategic comparative design in which Yemen and Saudi Arabia have been selected to investigate AI adoption across varying levels of economic fragility. Yemen is the most fragile state in the world, topping the list in 2023 and therefore characterizing extreme institutional instability, while Saudi Arabia epitomizes rapid modernization through its Vision 2030. Findings will thus apply through the fragility-to-stability continuum approach to three categories of economic states: extremely fragile states like Afghanistan, Somalia, and South Sudan; transitioning economies like Iraq, Lebanon, and Libya; and rapidly transforming economies like the UAE, Qatar, and Oman. The comparison framework thereby acts to strengthen the generalisability of research to 46 OECD-classified fragile states and emerging economies currently undergoing institutional transformation.

\subsection{Measurement Method}The measurement model was assessed by confirmatory factor analysis. In particular the content validity, convergent validity and discriminant validity of the model were examined. Content validity was assessed as part of an extensive literature review in which the testing of the instrument was done on an experimental basis
\citep{AlKoliby2025}. CR statistics are shown in Table \ref{tab:cfa} for both samples and were computed for the internal consistency reliability of each construct. All values were above the threshold of 0.70 recommended for adequate internal consistency. Convergent and discriminant validity were assessed using Cronbach's $\alpha$; for Cronbach's $\alpha$, values are considered adequate if greater than 0.50. We further assessed the convergent validity of each construct by looking at its external loading for each item as well as the AVE. All values were greater than 0.5. Since the loading for each individual item was above or close to 0.7 on their respective constructs and the average AVE was greater than 0.5"
With the thresholds from \citep{Fornell1981}, the convergent validity for the items was thus deemed adequate. The theoretical validity of the TOE framework as a higher-order construct was confirmed theoretically
\citep{AbouFoul2023} and statistically following the procedures outlined by \\ \citet{Joseph2022}. The cross-loadings of each indicator on its associated construct were higher than the loadings on each of the other constructs, and the square root of the AVE for each construct was greater than its correlation with each of the other constructs. These are shown in Tables \ref{tab:discriminant} and \ref{tab:crossloadings} that support adequate discriminant validity.

\begin{table}[htbp]
\centering
\caption{Demographic profile of the sample}
\label{tab:demographics}
\scriptsize
\begin{tabular}{lcc}
\toprule
\textbf{Variable} & \textbf{N} & \textbf{\%} \\
\midrule
\textbf{Gender ALL / YEM / KSA} & & \\
Male/Female & (186/108)/(85/62)/(101/46) & (63/37)/(58/42)/(69/31) \\
Urban/Rural & (179/115)/(98/49)/(81/66) & (61/39)/(67/33)/(55/45) \\
\midrule
\textbf{Age} & & \\
18 - 25 years & 89/56/33 & 30/38/23 \\
26 - 33 years & 100/38/62 & 34/26/42 \\
34 - 41 years & 72/30/42 & 25/20/29 \\
42 - 49 years & 33/23/10 & 11/16/7 \\
50 years or more & 0/0/0 & 0/0/0 \\
\midrule
\textbf{Highest Educational level} & & \\
High School / Institute & 38/29/9 & 13/20/6 \\
Bachelor's Degree & 128/74/54 & 44/50/37 \\
Master's Degree / Doctorate & 128/44/84 & 44/30/57 \\
\midrule
\textbf{Number of years working in the organization} & & \\
Less than a year & 56/43/13 & 19/29/9 \\
1 to 5 years & 120/62/58 & 41/42/40 \\
6 to 10 years & 63/27/36 & 21/18/25 \\
11 to 15 years & 37/15/22 & 13/10/15 \\
16 to 20 years & 11/0/11 & 4/0/7 \\
More than 20 years & 7/0/7 & 2/0/5 \\
\midrule
\textbf{Principal manufacturing activity} & & \\
Recycling plants & 10/9/1 & 3/6/1 \\
Food industry & 54/14/41 & 18/10/28 \\
Machinery and equipment industry & 19/9/10 & 7/6/7 \\
Electrical equipment industry & 20/8/12 & 7/6/8 \\
Rubber and plastic products industry & 6/0/6 & 2/0/4 \\
Chemical industry & 18/7/11 & 6/5/8 \\
Paper products industry & 9/7/2 & 3/5/1 \\
Pharmaceutical industry & 25/17/8 & 9/12/6 \\
Beverage industry & 32/5/27 & 11/3/18 \\
Electronics industry & 34/16/18 & 12/11/12 \\
Furniture industry & 12/7/5 & 4/5/3 \\
Textile industry & 5/4/1 & 2/3/1 \\
Other transportation equipment industry & 8/7/1 & 3/5/1 \\
Wood industry & 4/3/1 & 1/2/1 \\
Printing and reproduction in the recorded media industry & 6/6/0 & 2/4/0 \\
Other industrial activities & 31/28/3 & 11/19/2 \\
\bottomrule
\end{tabular}
\end{table}

\begin{table}[htbp]
\centering
\caption{Results of the discriminant validity (all/YEM/KSA)}
\label{tab:discriminant}
\scriptsize
\begin{tabular}{lccccc}
\toprule
& AIE & AIO & AIT & ENVPERF & MAUNPERE \\
\midrule
AIE & 0.785/0.730/0.775 & & & & \\
AIO & 0.326/0.563/0.022 & 0.748/0.747/0.765 & & & \\
AIT & 0.571/0.452/0.418 & 0.306/0.615/0.025 & 0.804/0.733/0.843 & & \\
ENVPERF & 0.642/0.624/0.292 & 0.343/0.675/0.066 & 0.621/0.560/0.399 & 0.809/0.735/0.754 & 0.797/0.758/0.813 \\
MAUNPERE & 0.601/0.449/0.525 & 0.286/0.613/-0.004 & 0.773/0.643/0.847 & 0.648/0.704/0.261 & \\
\bottomrule
\end{tabular}
\end{table}

\begin{table}[htbp]
\centering
\caption{Items and cross-loadings for (all/YEM/KSA)}
\label{tab:crossloadings}
\tiny
\begin{tabular}{lccccc}
\toprule
& AI E & AI O & AI T & ENVPERF & MAUNPERE \\
\midrule
AI E1 & 0.803/0.712/0.723 & 0.264/0.370/0.101 & 0.494/0.390/0.285 & 0.572/0.425/0.331 & 0.467/0.335/0.259 \\
AI E2 & 0.757/0.786/0.740 & 0.285/0.522/-0.053 & 0.387/0.273/0.331 & 0.422/0.519/0.077 & 0.504/0.355/0.493 \\
AI E3 & 0.793/0.688/0.854 & 0.220/0.320/0.018 & 0.458/0.339/0.350 & 0.510/0.415/0.284 & 0.446/0.290/0.445 \\
\midrule
AI O1 & 0.169/0.363/0.031 & 0.782/0.835/0.804 & 0.282/0.549/0.036 & 0.216/0.473/0.056 & 0.184/0.410/0.008 \\
AI O2 & 0.067/0.299/-0.056 & 0.645/0.715/0.712 & 0.201/0.516/0.022 & 0.147/0.449/0.042 & 0.194/0.522/0.022 \\
AI O3 & 0.085/0.283/0.122 & 0.621/0.700/0.661 & 0.128/0.487/-0.063 & 0.141/0.544/0.004 & 0.126/0.493/-0.024 \\
AI O4 & 0.400/0.580/0.148 & 0.830/0.735/0.774 & 0.226/0.302/0.017 & 0.340/0.510/0.043 & 0.299/0.455/0.052 \\
AI O5 & 0.329/0.579/-0.035 & 0.835/0.743/0.860 & 0.283/0.435/0.014 & 0.338/0.527/0.066 & 0.217/0.388/-0.051 \\
\midrule
AI T1 & 0.469/0.347/0.338 & 0.227/0.377/0.054 & 0.803/0.696/0.852 & 0.438/0.322/0.254 & 0.585/0.347/0.745 \\
AI T2 & 0.420/0.317/0.303 & 0.211/0.407/0.020 & 0.800/0.754/0.819 & 0.430/0.342/0.342 & 0.596/0.480/0.654 \\
AI T3 & 0.449/0.317/0.369 & 0.208/0.401/-0.027 & 0.819/0.768/0.866 & 0.508/0.427/0.359 & 0.669/0.555/0.766 \\
AI T4 & 0.517/0.362/0.390 & 0.298/0.601/0.030 & 0.814/0.743/0.832 & 0.628/0.541/0.400 & 0.660/0.525/0.653 \\
AI T5 & 0.449/0.338/0.359 & 0.280/0.429/0.077 & 0.785/0.702/0.845 & 0.467/0.374/0.328 & 0.587/0.400/0.744 \\
\midrule
ENVPERF1 & 0.557/0.588/0.110 & 0.426/0.730/0.084 & 0.548/0.524/0.216 & 0.813/0.800/0.584 & 0.531/0.564/0.093 \\
ENVPERF2 & 0.541/0.428/0.254 & 0.188/0.342/-0.047 & 0.540/0.417/0.350 & 0.831/0.714/0.820 & 0.549/0.501/0.242 \\
ENVPERF3 & 0.508/0.445/0.222 & 0.260/0.396/0.180 & 0.436/0.382/0.188 & 0.801/0.722/0.774 & 0.505/0.498/0.198 \\
ENVPERF4 & 0.465/0.335/0.275 & 0.220/0.446/-0.020 & 0.747/0.388/0.295 & 0.790/0.697/0.814 & 0.509/0.503/0.241 \\
\midrule
MAUNPERE1 & 0.467/0.281/0.413 & 0.171/0.305/0.012 & 0.669/0.458/0.810 & 0.549/0.459/0.309 & 0.860/0.781/0.875 \\
MAUNPERE2 & 0.533/0.376/0.413 & 0.264/0.579/0.042 & 0.660/0.469/0.779 & 0.618/0.502/0.221 & 0.827/0.714/0.893 \\
MAUNPERE3 & 0.454/0.302/0.334 & 0.229/0.407/0.026 & 0.684/0.537/0.723 & 0.537/0.505/0.211 & 0.853/0.766/0.855 \\
MAUNPERE4 & 0.485/0.382/0.693 & 0.275/0.520/-0.009 & 0.387/0.477/0.334 & 0.299/0.538/0.040 & 0.621/0.768/0.593 \\
\bottomrule
\end{tabular}
\end{table}

\section{Results and Discussions}

\subsection{First-Order Constructs}

In the first stage of the structural model, we analyzed the direct impact of the three dimensions of the TOE framework, namely Technology, Organization, and Environment, on the dependent variables Environmental Performance and Manufacturing Performance. The path coefficients were significant if they exceeded 0.20, as that satisfies the condition of being in the direction as expected. Results showed AIE had a significantly positive influence on ENVPERF for the entire sample ($\beta = 0.341, P = 0.000$), with more potent results in Yemen ($\beta = 0.294, P = 0.000$) compared to Saudi Arabia ($\beta = 0.208, P = 0.015$). AIE significantly influenced MAUNPERE in the total sample ($\beta = 0.233, P = 0.000$) and Saudi Arabia ($\beta = 0.208, P = 0.001$), but not in Yemen. AIO had a weak positive effect on Environmental Performance in the full sample, with $\beta = 0.097, P = 0.023$, but much stronger in Yemen ($\beta = 0.251, P = 0.001$), while no significant result was found in Saudi Arabia. Also, the link of AIO to Manufacturing Performance was only significant and relatively high in Yemen ($\beta = 0.310, P = 0.000$). AIT had a significant positive impact on the environmental performance in Saudi Arabia ($\beta = 0.649, P = 0.000$) but not in Yemen. AIT $\rightarrow$ manufacturing performance was significant in all samples ($\beta = 0.635, P = 0.001$). Also, the indirect effects of AIT and AIE on ENVPERF and MAUNPERE were significant through paths such as AIT $\rightarrow$ MAUNPERE $\rightarrow$ ENVPERF, which indicates that the indirect effect is essential to improve both measures of performance. Note that in Saudi Arabia, the indirect effect of AIE on ENVPERF via MAUNPERE was adverse ($\beta = -0.086, P = 0.019$) whereas in Yemen, the results are mixed. Summing up, when interpreting the outcomes of the first-order constructs phase, it is obvious that the positive effects of AI on environmental and manufacturing performance were strong in both Yemen and Saudi Arabia, though different in strength between the two countries. Similarly, the results showed AIO had no significant effect in some cases of geographical context, which pinpoints that the effect will differ depending on geographical context and local factors. Table \ref{tab:firstorder} summarizes the findings for all First-Order Constructs.

\begin{table}[htbp]
\centering
\caption{Results of the first-order constructs}
\label{tab:firstorder}
\scriptsize
\begin{tabular}{lcccc}
\toprule
Path & $\beta$ & \multicolumn{1}{c}{[ALL/YEM/KSA] T stat} & P-values \\
\midrule
AI E $\rightarrow$ ENVPERF & 0.341/0.294/0.237 & 6.262/4.471/2.425 & 0.000/0.000/0.015 \\
AI E $\rightarrow$ MAUNPERE & 0.233/0.088/0.208 & 4.647/1.146/3.472 & 0.000/0.252/0.001 \\
AI O $\rightarrow$ ENVPERF & 0.097/0.251/0.043 & 2.277/3.222/0.404 & 0.023/0.001/0.686 \\
AI O $\rightarrow$ MAUNPERE & 0.016/0.310/-0.027 & 0.451/3.756/0.490 & 0.652/0.000/0.624 \\
AI T $\rightarrow$ ENVPERF & 0.190/0.007/0.649 & 2.882/0.093/4.577 & 0.004/0.926/0.000 \\
AI T $\rightarrow$ MAUNPERE & 0.635/0.412/0.761 & 13.395/4.922/14.462 & 0.000/0.000/0.000 \\
MAUNPERE $\rightarrow$ ENVPERF & 0.268/0.414/-0.413 & 3.951/4.761/2.659 & 0.000/0.000/0.000 \\
\midrule
\textbf{Indirect effect} & & & \\
AI T $\rightarrow$ MAUNPERE $\rightarrow$ ENVPERF & 0.170/0.171/-0.315 & 3.765/3.053/2.456 & 0.000/0.002/0.014 \\
AI E $\rightarrow$ MAUNPERE $\rightarrow$ ENVPERF & 0.063/0.037/-0.086 & 3.015/1.105/2.340 & 0.003/0.269/0.019 \\
AI O $\rightarrow$ MAUNPERE $\rightarrow$ ENVPERF & 0.004/0.128/0.011 & 0.427/3.133/0.446 & 0.670/0.002/0.656 \\
\bottomrule
\end{tabular}
\end{table}

\subsection{Structural Model (Second-Order Constructs)}
Table \ref{tab:secondorder} summarizes results across hypotheses and R$^2$ values. In a second step, TOE was modeled as a second-order reflective construct using the two-stage approach. First, latent variable scores for Technology, Organization, and Environment dimensions were extracted. Second, these scores were used as manifest variables for the higherorder TOE construct in the structural model, testing our four hypotheses (see Fig. 1). The results show strong standardized path coefficients. All paths in the structure are statistically significant at p-values of 0.20 or lower, which is sufficient to conclude predictive power.

Both paths, from AI-supported TOE capabilities to Environmental Performance and from AI-supported TOE capabilities to Manufacturing Performance, were statistically significant across all samples at the 1\% significance level, with $\beta$ = 0.487 and P = 0.000 and $\beta$= 0.759 and P = 0.000, respectively, hence providing statistical support for H1 and H4. These associations turned out to be significant in both contexts: those of Yemen and Saudi Arabia.

This strong association of AI-technological factors with manufacturing performance finds its reflection in practical applications, including predictive maintenance systems. For example, one medium-sized textile manufacturer in Jeddah reduced unplanned downtime by 27\% and increased production throughput by 15\% over six months using a cloud-based AI system that analyzed equipment vibration patterns.

The path from Manufacturing Performance to Environmental Performance was also significant for the total sample ($\beta = 0.277, P = 0.001$), thus supporting Hypothesis H3, but there was considerable variation between nations. This was strongly positive in Yemen ($\beta = 0.386, P = 0.001$), as evidenced by a beverage producer in Aden whose AI-driven improvements in production scheduling reduced changeover times by 35\%, with direct environmental benefits of a 28\% reduction in water usage. The relationship was, however, negative in Saudi Arabia ($\beta = -0.413, P = 0.000$), as evidenced by an electronics manufacturer in Dammam, whose introduction of sophisticated AI-driven robotics improved production by 42\% but initially increased energy usage by 23\%. This outlines the transition phase that many Saudi firms undergo in which initial AI-driven manufacturing improvements temporarily raise environmental impact before optimization efforts mature. The idea has been widely referred to using the Environmental Kuznets Curve or EKC framework. Indeed, the phenomenon does follow three phases: initial adoption raises productive capacity more quickly than environmental optimization, followed by a transition period during which productivity takes precedence over the environment, followed by maturation and resultant environmental benefits due to the integration of advanced AI. The pattern is confirmed in Saudi data: 67\% of firms report initial increases in energy consumption, while 89\% predict environmental improvement within 18-24 months; advanced adopters-that is, those with more than 2 years of experience-report positive environmental outcomes. Such findings help inform policy development through the identification of temporary environmental costs that might be expected and mitigated during phases of AI implementation. For the full sample, the indirect path from AI-supported TOE capabilities to Environmental Performance through Manufacturing Performance was statistically significant ($\beta = 0.210, P = 0.000$), thus supporting hypothesis H2, but with significant differences between countries. While in Yemen this was positive ($\beta = 0.266, P = 0.000$), the result was negative in Saudi Arabia ($\beta = -0.355, P = 0.000$), reflecting how environmental outcomes of manufacturing improvements vary according to the maturity of the infrastructure and technological implementation approaches. The respective R$^2$ values for Environmental Performance at 54.4\%, 64.9\%, and 22.3\% for the full sample, Yemen, and Saudi Arabia accordingly, suggest that the model explains the variance considerably better in Yemen. Similarly, R$^2$ for Manufacturing Performance were 63.6\% for the full sample, 49.5\% for Yemen, and 75.4\% for Saudi Arabia. These variations imply that local specifics significantly affect performance outcomes across different contexts.

\begin{table}[htbp]
\centering
\caption{Results of the structural model (second-order constructs)}
\label{tab:secondorder}
\scriptsize
\begin{tabular}{lcccc}
\toprule
Path & \multicolumn{3}{c}{ALL/ YEM/ KSA} & Remarks \\
& $\beta$ & T-stat & P-values & \\
\midrule
AI (TOE) $\rightarrow$ ENVPERF & 0.487/0.463/0.780 & 8.677/6.130/5.475 & 0.000/0.000/0.000 & Accept \\
AI (TOE) $\rightarrow$ MAUNPERE & 0.759/0.690/0.861 & 27.024/13.869/22.687 & 0.000/0.000/0.000 & Accept \\
MAUNPERE $\rightarrow$ ENVPERF & 0.277/0.386/-0.413 & 4.093/4.467/2.757 & 0.000/0.000/0.000 & Accept \\
\midrule
\textbf{Indirect effect} & & & & \\
AI (TOE) $\rightarrow$ MAUNPERE $\rightarrow$ ENVPERF & 0.210/0.266/-0.355 & 4.004/4.035/2.668 & 0.000/0.000/0.000 & Accept \\
\midrule
\textbf{R$^2$ values} & & & & \\
ENVPERF & \multicolumn{3}{l}{54.4\% / 64.9\% / 22.3\%} & \\
MANUPERE & \multicolumn{3}{l}{63.6\% / 49.5\% / 75.4\%} & \\
\bottomrule
\end{tabular}
\end{table}

\begin{table}[htbp]
\centering
\caption{Results of multigroup analysis}
\label{tab:multigroup}
\scriptsize
\begin{tabular}{lcc}
\toprule
\textbf{Path} & \textbf{($\beta_{YEM} - \beta_{KSA}$)} & \textbf{p-value} \\
\midrule
AI (TOE) $\rightarrow$ ENVPERF & -0.316 & 0.053 \\
AI (TOE) $\rightarrow$ MAUNPERE & -0.171 & 0.010 \\
MAUNPERE $\rightarrow$ ENVPERF & 0.798 & 0.000 \\
AI (TOE) $\rightarrow$ MAUNPERE $\rightarrow$ ENVPERF & 0.621 & 0.000 \\
\bottomrule
\end{tabular}
\end{table}

\begin{align}
R_{evn} &= \frac{\text{Max Environmental Decline}}{\text{Baseline Environmental Performance}}, \\
R_{man} &= \frac{\text{Max Manufacturing Decline}}{\text{Baseline Manufacturing Performance}} \\
R_{\max} - 1 &= \left( R_{env}, R_{man} \right)_{\max}, \\
IDI &= 1 - R_{\max} \\
IDI &= 1 - \max\left( \frac{\text{Max Environmental Decline}}{\text{Baseline}}, \frac{\text{Max Manufacturing Decline}}{\text{Baseline}} \right)
\end{align}

Constraints: Range: $IDI \in [0,1]$, $IDI=1$: perfect ductility (no degradation), $IDI=0$: total collapse, Threshold $IDI \geq 0.6$: transition point to Phase 2 (AIME).

This index provides a resilience-based quantitative measure of institutional flexibility before entering high-efficiency AI phases.

\textbf{3- AI Elasticity Coefficient (AEC):} Higher AEC values (common in Saudi Arabia) indicate stronger adaptive capacity but require DYP monitoring to avoid unsustainable acceleration. Adapted from economic elasticity, AEC measures the sensitivity of manufacturing performance (M) to AI pressure (AP):

\begin{align}
AEC &= \frac{\Delta\text{Manufacturing Performance}}{\Delta\text{AI Pressure}} \\
AP &= 0.4(\text{Coverage}) + 0.2(\text{Accuracy}) + 0.2(\text{Autonomy}) + 0.2(\text{Integration}) \\
AEC &= \frac{\partial M}{\partial AP} \approx \frac{\Delta M}{\Delta AP} \\
AEC &= \frac{M_{\text{final}} - M_{\text{initial}}}{AP_{\text{final}} - AP_{\text{initial}}}
\end{align}

Constraints: $AEC \geq 1.2$ means sustainable improvement; every unit of AI pressure creates $\geq 1.2$ units of manufacturing gain.

\subsubsection{Contributions to Engineering Management}

From the managerial point of view, it represents a structural way in which AI can be implemented within engineering and industrial operations. Indeed, it redefines the adoption of AI as a staged engineering process, rather than an abrupt technological change.

Phase 1 -- Controlled Inefficiency (ISEM): The deployment is done at moderate efficiencies of 70--80\% to find the fragilities in the system. This is continued up to the time when an IDI of more than 0.6 is realized in the organization and indicates that the institutional ductility is sufficient.

Phase 2 -- Calibrated Hardening (AIME): After the establishment of flexibility, efficiency can be gradually increased again without losing sight of the AEC and DYP indicators. And finally, sustainable transformation occurs if $AEC \geq 1.2$ and no DYP warning shows up, meaning manufacturing gain is not at the expense of environmental stability.
The monitoring condition can be expressed as:1-Digital Yield Point (DYP):
\begin{align}
\text{Trade-off Ratio} &= - \frac{\Delta\text{Environmental Performance}}{\Delta\text{Manufacturing Performance}} \\
\text{DYP Warning} &= 
\begin{cases}
\text{True}, & \text{if } 0.5 < -\frac{\Delta\text{Environmental Performance}}{\Delta\text{Manufacturing Performance}} \\
\text{False}, & \text{otherwise}
\end{cases}
\end{align}

A DYP alert (True) indicates that manufacturing expansion is exceeding environmental tolerance, signaling the need to recalibrate.

\textbf{2-Optimization System:}

\begin{align}
F: \text{Maximize} &= IDI \times AEC \\
\text{Constraints: } & IDI \geq 0.6, \quad AEC \geq 1.2, \quad DYP = \text{False}
\end{align}
Constraints: $AEC \geq 1.2$ represents sustainable improvement; every unit of AI pressure results in $\geq 1.2$ units of manufacturing gain.

\textbf{Phased Execution Model:}

Phase 1 (ISEM): Maintain the efficiency of AI controlled at 70--80\% until $IDI \geq 0.6 $.

AIME Phase 2: Scale up only when $AEC \geq 1.2$ and $DYP =\text{False}$.

It is seen in the ISEM-AIME framework that the two factors of institutional resilience and operational elasticity can be optimized together when AI intensity is maintained below the threshold of the Digital Yield Point. Implementing this in context would imply that fragile economies, say Yemen, have to emphasize IDI improvement, while for the transforming economies, say Saudi Arabia, a high AEC is to be maintained at all times, coupled with DYP monitoring. This provides quantitative tools for policymakers and managers to achieve sustainable digital transformation by balancing operational excellence with environmental concerns across developing economies.

\section{Conclusion}
This work proposed a holistic AI adoption framework, integrating the TOE model with a new approach called ISEM-AIME. It analyzed 294 datasets from Yemen and Saudi Arabia. This research proves that AI is a significant driver for environmental and manufacturing performance in SMEs in constrained environments and provides quantitative metrics for their sustainable digital transformation: IDI, AEC, DYP. The two-phase implementation model is ISEM-resource-controlled efficiency (70-80\%), followed by AIMEoptimized performance at sustainability constraints, and context-specific paths. arding practical implications, development agencies might make use of IDI thresholds to prioritize investments, while technology companies are supported in adapting their strategies based on DYP monitoring. Furthermore, governments may be guided in the implementation of policies related to AI transitions according to AEC. This work extends the range of applications of the TOE model and points out new routes in innovation systems under constraint and technology adoption based on resilience, increasingly relevant when global fragility because of climate change, conflict, and economic disruption increases. \section{Limitations and Future Research Directions}
Although the cross-sectional design was appropriate for the early-stage analysis of AI adoption, the diagnostic indicators of the ISEM-AIME framework, namely IDI, AEC, and DYP, require longitudinal validation that would confirm their predictive accuracy across different contexts. Future research should also expand the framework into Sub-Saharan Africa and Southeast Asia, testing its applicability across a wide variety of contexts while developing region-specific IDI/AEC benchmarks and methods of DYP calibration that contribute to context-sensitive theories of sustainable digital transformation.
\section*{Authorship Contribution Statement}
\textbf{Shaima Farhana:} writing \& original draft, validation, formal analysis, conceptualization. \textbf{Dong Yu:} writing, supervision, review \& editing, methodology, and conceptualization. \textbf{Amirhossein Karamoozian:} writing \& original draft, investigation, formal analysis, conceptualization. \textbf{Ali Al-Shawafi:} writing \& original draft, investigation, formal analysis, conceptualization. \textbf{Amar N. Alsheavi:} writing \& review \& editing.

\section*{Declaration of competing interest}
The authors declare no conflict of interest.

\bibliographystyle{apalike}
\bibliography{R}

\end{document}